# Finding the Ion Current Density of Microtubules by defining a potential function for the same and Solving the time independent Schrodinger Equation.

*Shantanav Chakraborty, Naman Joshi,  Anop Singh, Ety Mittal*

*Indian Institute of Technology, Rajasthan*

**Abstract:** In this paper, we model a microtubule based on its dimer resolution structure. First, the fundamental structural components were studied and then electrostatic potential function for a single monomer was calculated. Subsequently, the potential function inside a single monomer due to a ring of dimers was obtained. Considering the potential due to protofilament-protofilament interaction with a monomer in a B crystal structure of a microtubule, we obtain a double well potential wall. Quantum mechanically the ions can pass through this wall because of the Tunnelling effect. We solve the time independent Schrodinger Equation, calculate the transmission efficiency of ion flow and use the latter in the calculation of ion current density.

**Keywords**:   Microtubules,   Potential   Wall,   Tunnelling,    Quantum   mechanics

## 1. Introduction:

Microtubules (1, 2, 3) (MTs) are protein filaments of the cytoskeleton with their outer diameter roughly 23 nm, and a hollow interior with a diameter of roughly 15 nm. Their lengths vary but commonly reach 5-10 microm dimensions. They are composed of 13 protofilaments .These protofilaments are chains of $\alpha - \beta$ dimers. MTs are found in nearly all eukaryotic cells and they perform a variety of key cellular functions. In addition, to providing rigidity and structural integrity to a living cell, they serve as tracks for motor protein transport. They also form the core of cilia and flagella which beat in a coordinated manner to either move objects along the cell membrane or to propel the cell through its environment.

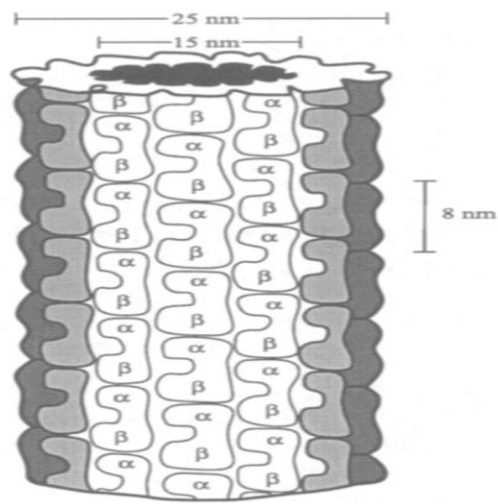

*Fig 1. A section of a typical microtubule demonstrating the helical nature of its construction and the hollow interior which is filled with cytoplasm. Each vertical column is known as a protofilament and the typical MT has 13 protofilaments.*

The general structure of MTs has been well established experimentally (2). A small difference between the $\alpha$ and $\beta$ monomers of tubulin allows the existence of several lattice types (shown in Fig 2). Moving around the MT in a left-handed sense, protofilaments of the A lattice have a vertical shift of 4.92 nm upwards relative to their neighbours. In the B lattice this offset is only 0.92 nm because the $\alpha$ and $\beta$ monomers have switched positions in alternating filaments. This change results in the development of a structural discontinuity in the B lattice known as a seam (3).

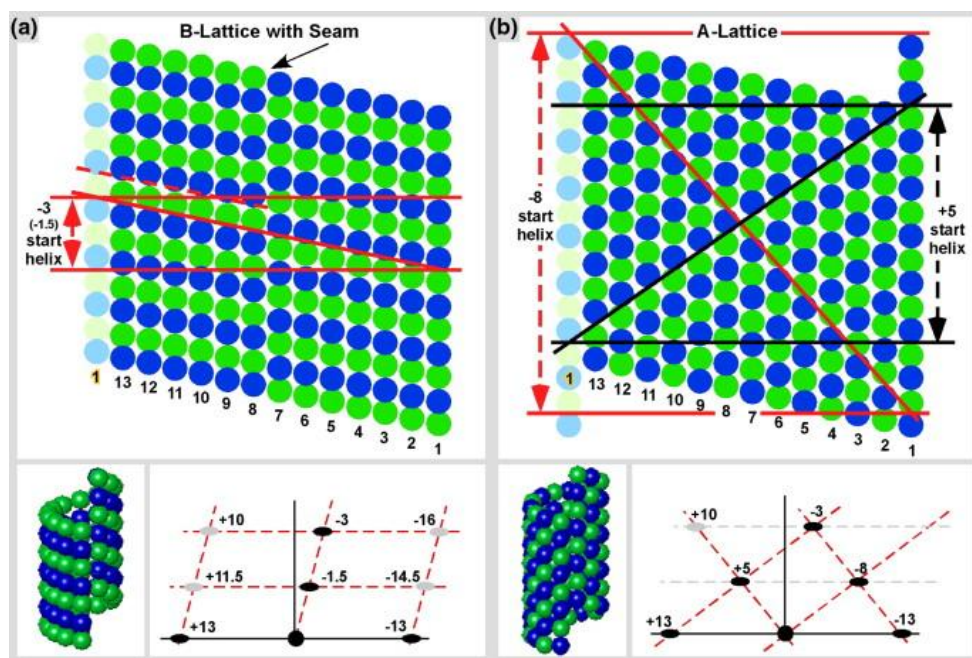

*Fig 2. The 13A and 13B MT lattices. In the A lattice, perfect helical spirals are formed and in the B lattice, there is a structural discontinuity known as the seam.*

## 2. Methods

### a. Electrostatic Potential Around a Tubulin

The method of distributed multi pole analysis (DMA) provides a fairly accurate means of calculating the electrostatic field around a biomolecule. Diatomics, triatomics, and tetratomics are described to high precision with the use of monopoles, dipoles and quadrupoles providing an accurate picture of molecular bonding. In the case of $\alpha$ and $\beta$ tubulin, each monomer is comprised of approximately 450 amino acids, i.e., on the order of 7000 atoms. We find the electrostatic potential on a monomer. A detailed examination shows that, each monomer is formed by a core of two $\alpha$ and $\beta$ sheets (4) that are surrounded by $\alpha$-helices. Having obtained the charge distribution on the tubulin surface, we have attempted to investigate its role in the microtubule lattice formation and its interaction with ions and macromolecules. We have confined our examination largely to the surfaces of tubuIin that form the exterior surface and the protofilament-protofilament contacts when assembled into a MT. The profiles of the electrostatic potential (2, 5) are shown in Fig 3.

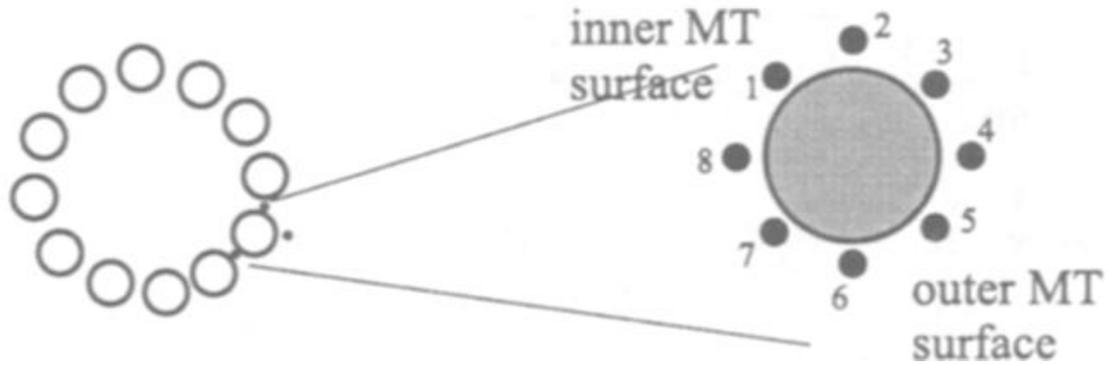

Fig 3. A MT cross-section illustrates where the electrostatic potential was examined *along lines parallel to the protofilament axis (perpendicular to the page) in the preceding figure.*

## b. Formula for Calculation of Electrostatic Potential on a monomer due to a ring of $\alpha - \beta$ dimers

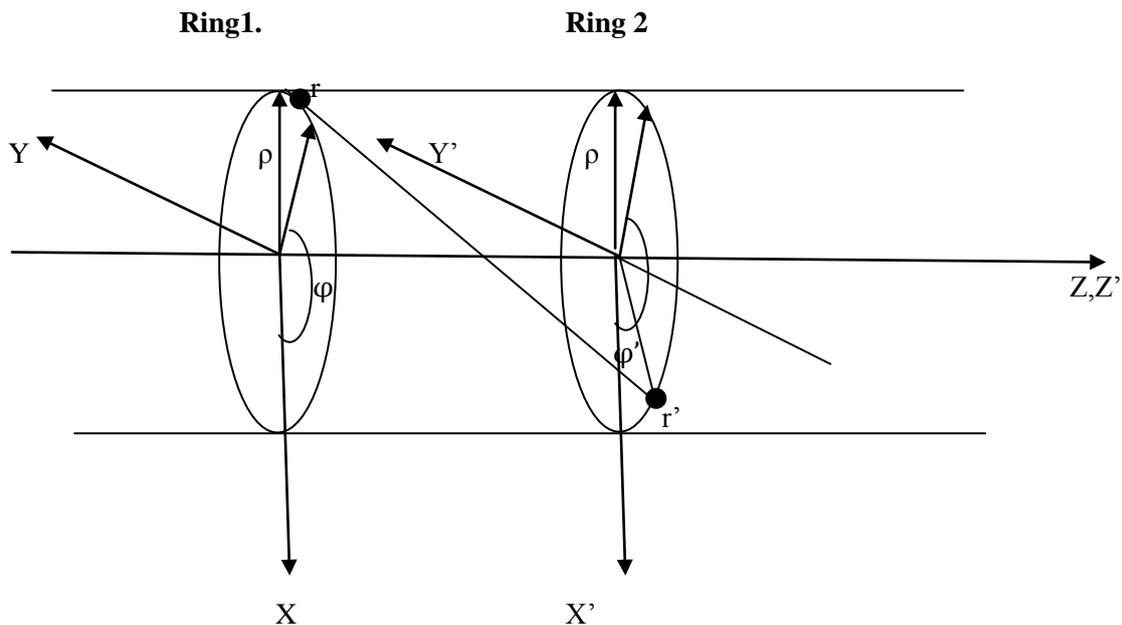

*Fig 4. The potential at a monomer r' situated at a ring2 due to a monomer r at Ring 1. Let us assume that the monomer r' is situated at a distance of z from the monomer r in the Z-axis direction, The radius of both the rings of microtubules are = $\rho$=7.5nm*

For a microtubule, each monomer will experience a potential from a monomer within a ring of dimmers as well as a monomer situated at subsequent rings throughout the microtubule. In this section, we calculate the potential on a monomer due to another monomer situated in another ring.

Now, charge of a monomer $\alpha$ = Charge of $\beta$ = $e$ = $1.6 \times 10^{-19} C$

As there are 13 $\alpha$ and $\beta$ monomers in a ring,

Linear Charge Density of a Ring = $\lambda = \dfrac{13e}{2\pi\rho}$

Radius of both rings = $\rho$
Charge on arc $d\theta$ $(dq) = \lambda.\rho.d\theta$
Azimuthal Angle varies from 0 to $2\pi$

Electrostatic Potential due to element = $dV = k.\dfrac{dq}{|r-r'|} = k.\dfrac{\lambda.\rho.d\theta}{|r-r'|}$,

where $k = \dfrac{1}{4\pi\epsilon\epsilon_0}$, $\epsilon$ = Permittivity of Water

Now, clearly from Fig 4, $|r-r'| = \sqrt{(x-x')^2+(y-y')^2+(z-z')^2}$

$$\text{Electrostatic Potential } V = \int_0^{2\pi} \dfrac{k.13.e.d\theta}{\sqrt{(x-x')^2+(y-y')^2+(z-z')^2}}$$

$$= \int_0^{2\pi} \dfrac{k.13.e.d\theta}{\sqrt{(\rho\cos\theta - \rho\cos\theta')^2 + (\rho\sin\theta - \rho\sin\theta')^2 + (z-0)^2}}$$

$$= \int_0^{2\pi} \dfrac{k.13.e.d\theta}{\sqrt{2\rho^2(1-\cos(\theta-\theta')) + z^2}}$$

Let, $\theta - \theta' = \varphi$
$\Rightarrow d\theta = d\varphi$

$$V(\text{in eV}) = \dfrac{k}{\sqrt{2}\rho} \int_{-\theta'}^{2\pi-\theta'} \dfrac{d\varphi}{\sqrt{N^2 - \cos\varphi}}$$

## 3. Results

### a. Potential on a monomer

This value of Potential varies according to $\theta'$ and z. If we look at Fig 4, we can observe that by putting z=0, we will obtain the values of potential on one monomer due to another. Again, by varying both $\theta'$ and z, we can obtain the potential on a monomer due to any other monomer located in any other ring.

We have taken the help of SCILAB, to calculate the necessary potentials. First, we calculated the potential on a monomer due to a monomer at an adjacent ring for $\theta' = 0$. On varying the value of z, we obtain the total potential on a monomer due to monomers situated at similar positions in the entire microtubule.

It is worth mentioning, that we have assumed that the length of the microtubule is large enough and as the potential $V \alpha \frac{1}{z}$, the potential gradually decreases and the effect of potential on a monomer due to rings situated far away from it, is negligible.

In SCILAB, we used Simpson's 3/8 rule to find out the value of the integral at different points on the microtubule. Firstly, we used a loop for the variable z and varied it from 0 to 5microm at intervals of 4nm. We have varied $\theta'$ from 0 to 360° at intervals of 27.7°(All degrees have been converted into radians of course). For each such value of z and $\theta'$, we obtain a potential value. Ultimately we sum all such potentials to obtain the potential on a monomer due to all other monomers.

Naturally, a question may arise as to why $\theta'$ was varied at intervals of 27.7°. Through Fig 5, we clearly specify the reason. The electrostatic potential on one monomer due to another within a ring of the microtubule can be obtained for values when z=0. Clearly, for z=0, and for different values of $\theta'$, we obtain the required potential on the monomer on a ring due to other monomers in the same ring.

Now, each ring contains 13 monomers with each monomer situated at an angular displacement of $\frac{360°}{13} = 27.7°$ from each other.

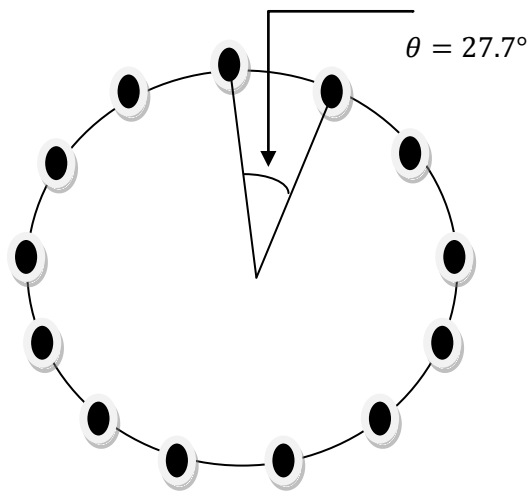

*Fig 5. One Ring of a microtubule containing 13 $\alpha - \beta$ dimers*

Clearly the angular displacement between any two monomers separated by n monomers
$= n\theta = 27.7.n$

**The value of potential obtained (V) on a monomer = 0.21375eV**.

**b. Solution of the time independent Schrodinger Equation and finding out the Transmission Probability for different ions through the microtubule**

We have calculated the electrostatic potential inside $\alpha$ (or $\beta$) monomer which in our case are equal. The contact surface between $\alpha$ and $\beta$ monomer experiences a high potential value which in turn provides a potential barrier to the ions which are flowing from one dimer to another. The value of such barrier potential is relatively as compare to the potential value inside $\alpha$ and $\beta$ monomer (1), and the ion flow can be described with Tunnelling Effect.

The value of potential wall has been experimentally verified (1, 2, 7) and used in our calculation. In this paper, we take value of potential well equal to the potential inside $\alpha$ monomer and apply Schrodinger Wave function to calculate transmission probability (T). Transmission Probability describes the possibility of an ion to cross the potential wall without having sufficient energy. Generally, this value is small but if the ions striking with potential wall is high, the more ions will tunnel through the barrier resulting in high tunnelling current density

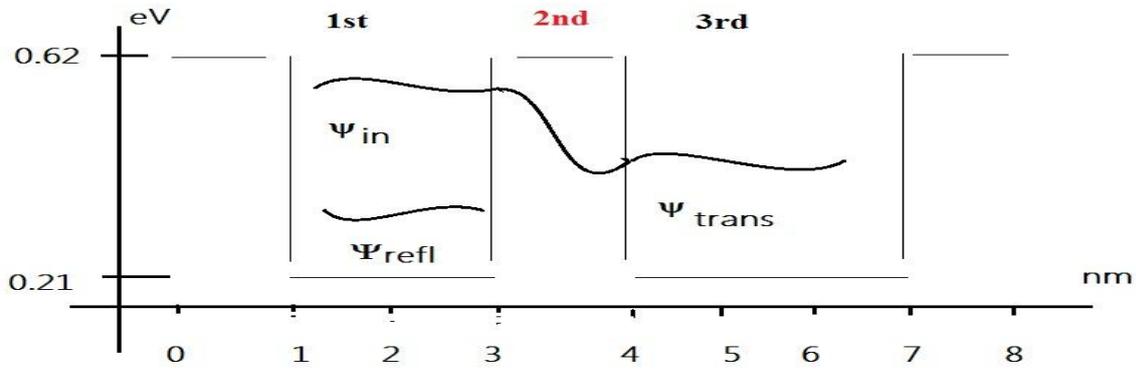

*Fig 6: The wave function distribution for a $\alpha - \beta$ dimer.*

Since the heterodimer structure has 8 nanometer length the partition of $\alpha$ monomer, length of potential well and length of $\beta$ monomer can be taken as 2, 1 and 3 nm respectively. The approximate function of electrostatic potential experienced by ions is shown in Fig. 6 as a function of distance. The time independent Schrodinger Equation for cylindrical co-ordinates (6) is given by:

$$\left(\frac{d^2}{d\rho^2} + \frac{1}{\rho}\frac{d}{d\rho} + \frac{1}{\rho^2}\frac{d^2}{d\theta^2} + \frac{d^2}{dz^2}\right)\psi + \frac{2m(E-V)}{\hbar^2}\psi = 0$$

Since the ion flow is expected only in the z direction the terms corresponding to $\theta$ and $\rho$ vanish and reduced Schrodinger equation is obtained as follows:

$$\frac{d^2}{dz^2}\psi + \frac{2m(E-V)}{\hbar^2}\psi = 0$$

Where E= total energy of ion and V= potential function. We define two potential level as p1=0.2217eV and p2=0.62eV. (1, 2, 4, 5, 9, 10)

Wave function for first region can be written as

$$\frac{d^2}{dz^2}\psi + \frac{2m(E)}{\hbar^2}\psi = 0,$$ where E= p1 eV.

So solution of this differential equation can be obtained as follows

ψ=A cos (k1z) + B sin(k1z)

where $k1 = \sqrt{\frac{2mp1}{\hbar^2}}$

For the second region, Schrodinger equation will have following format:

$$\frac{d^2}{dz^2}\psi + \frac{2m(E-V)}{\hbar^2}\psi = 0$$

where E= p1 and V=p2 (potential into 2$^{nd}$ region) and since E<V we have following solution of the above equation:

ψ=$Ce^{k2z} + De^{-k2z}$

Where $k2 = \sqrt{\frac{2m(p2-p1)}{\hbar^2}}$

Thus, Transmission probability is given by

$$T = \sqrt{\frac{16}{4+\frac{k1^2}{k2^2}}} e^{-2k2d}$$

Where d is the length of potential barrier and,

$k1 = \sqrt{\frac{2mp1}{\hbar^2}}$

$k2 = \sqrt{\frac{2m(p2-p1)}{\hbar^2}}$

d = 1 nm

Through the protofilaments of the microtubules several types of ions flow. The ion current density as well as probability of Transmission varies accordingly depending on the. This will depend on the mass as well as the the mobility of the ions through the microtubule protofilament.

The ion current density equation is given as $J = nqv_\gamma T$, where T = Transmission Probability of the ion and $v_\gamma$ = Diffusion velocity of the ion. For different ions, the values of J and T are given in Table I below.

| ION | VELOCITY | TRANSMISSION PROBABILITY | CURRENT DENSITY |
|---|---|---|---|
| Mg2+ | 0.6 micron/s | $1.318 \times 10^{-98}$ | $2.53 \times 10^{-123}$ |
| Ca2+ | 0.26 micron/s | $1.58 \times 10^{-126}$ | $1.31 \times 10^{-151}$ |
| Sr2+ | 0.025 micron/s | $1.37 \times 10^{-186}$ | $1.09 \times 10^{-212}$ |
| Ba2+ | 0.014 micron/s | $2.06 \times 10^{-233}$ | $9.22 \times 10^{-260}$ |

*Table I: T and J values for various ions flowing across the protofilament of a MT.*

## 4. Conclusion

We have obtained the potential profile for a monomer due to every other monomer throughout the microtubule. Using this value of potential on a monomer and the using the value of the potential wall between the $\alpha - \beta$ dimers, we solve the time independent Schrodinger Equation and subsequently obtain the transmission probability for tunnelling of various ions through the protofilament. For each such ion we again calculate the ion current density. This work, to the best knowledge of the authors, is the first of its kind and shall help in further simplifying the functions of microtubules. Here, the authors have assumed the size of both $\alpha$ and $\beta$ dimers to be the same. In future, this assumption may be gotten rid of and further accuracy in results may be obtained.